\documentclass{ws-ijmpa}
\usepackage{amssymb,graphicx,amsmath,graphics,subfigure}
\usepackage[super,compress]{cite}
\begin{document}

\title{Scattering solutions of the Klein-Gordon equation for a step potential with hyperbolic tangent potential}

\author{Clara Rojas\footnote{clararoj@gmail.com}}
\address{CEIF IVIC Apdo 21827, Caracas 1020A, Venezuela}

\maketitle

\begin{history}
\received{28 March 2014}
\end{history}

\begin{abstract}
We solve the Klein-Gordon equation for a step potential with hyperbolic tangent potential. The scattering solutions are derived in terms of hypergeometric functions. The reflection coefficient $R$ and transmission  coefficient $T$ are calculated, we observed superradiance and transmission resonances.

\end{abstract}

\ccode{PACS numbers: 02.30.Gp,03.65.Pm, 03.65.Nk}

%Introduction
\section{Introduction}
\label{Introduction}

The study of the scattering solutions of the Klein-Gordon equation \cite{rojas:2005,rojas:2006a,rojas:2006b,rojas:2007,arda:2011,alpdogan:2013 } and for the Dirac equation \cite{kennedy:2002,villalba:2003} with different potentials has been extensively studied in recent years.

The  phenomenon of superradiance, when the reflection coefficient $R$ is greater than one,  has been widely discussed. Manogue \cite{manogue:1988} discussed the superradiance on a potential barrier for Dirac and Klein-Gordon equations. Sauter \cite{sauter:1931b} and Cheng \cite{cheng:2009} have studied the same phenomenon for the hyperbolic tangent potential with the Dirac equation.  Superradiance for the Klein-Gordon equation with this particular potential has been studied for Cheng\cite{cheng:2009} and Rojas\cite{rojas:2014}. 

Transmision resonances, when the transmission coefficient $T$ is one, has been observed in the scattering of scalar relativistic particles in Klein-Gordon equation \cite{rojas:2005,rojas:2007} and the Dirac equation \cite{kennedy:2002,villalba:2003}.

In this paper we have calculated the scattering solutions of the Klein-Gordon equation in terms of hypergeometric functions for a step potential with hyperbolic tangent potential. The reflection coefficient $R$ and transmission coefficient $T$ are calculated numerically. The behaviour of the reflection $R$ and transmission $T$ coefficients is studied for three different regions of energy. We have observed superradiance\cite{cheng:2009, wagner:2010} and transmission resonances.

This paper is organized of the following way. Section \ref{Klein-Gordon} shows the one-dimensional Klein-Gordon equation. In section \ref{btanh} the step potential with hyperbolic tangent potential is shown. Section \ref{scattering} shows the scattering solutions and the behaviour of the reflection coefficient $R$ and transmission coefficient $T$. Finally, in section \ref{conclusion} conclusions are discussed.

%Klein-Gordon
\section{The Klein-Gordon equation}
\label{Klein-Gordon}

The one-dimensional Klein-Gordon equation to solve is, in natural units $\hbar=c=1$ \cite{greiner:1987}

\vspace{-0.5cm}
\begin{equation}
\label{klein}
\frac{d^2\phi(x)}{dx^2}+\left\{\left[E-V(x) \right]^2-m^2 \right\}\phi(x)=0,
\end{equation}
where $E$ is the energy, $V(x)$ is the potential and, $m$ is the mass of the particle.

%Step + Hyperbolic tangent
\section{Step potential with hyperbolic tangent potential}
\label{btanh}

\begin{figure}[htbp]
\begin{center}
\includegraphics[scale=0.50]{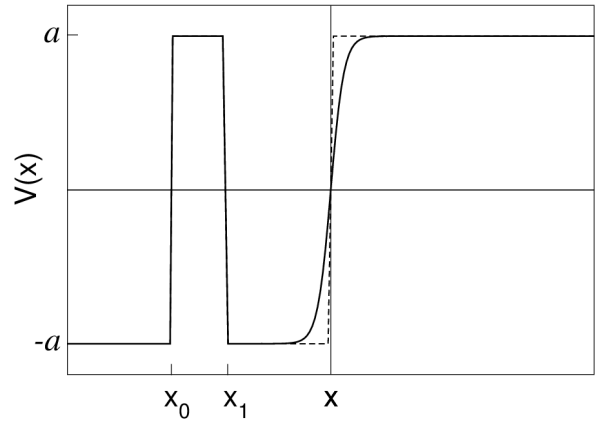}
\end{center}
\caption{\label{fig_pot}{Step potential with hyperbolic tangent potential for $a=5$ with $b=2$ (solid line) and $b=50$ (dotted line).}}
\end{figure}

The step potential with hyperbolic tangent potential has the following form

\begin{equation}
\label{potential}
V(x) =
\left\{
	\begin{array}{ll}
	       -a, \quad \textnormal{for} \quad x<x_0,\\
	\quad	a, \quad \textnormal{for} \quad x_0<x<x_1,\\
	\quad	a\,\tanh(b\,x), \quad \textnormal{for} \quad x>x_1,\\
	\end{array}
\right.
\end{equation}
where $a$ represents the height of the potential and $b$ gives the smoothness of the curve. The value of $x_0$ and $x_1$ represents the position of the step. The form of step potential with hyperbolic tangent potential is showed in the Fig. (\ref{fig_pot}). From Fig. \ref{fig_pot} we can note that the step potential with hyperbolic tangent potential reduces to two barriers or two wells for $b \rightarrow \infty$. 

%Scattering States
\section{Scattering States}
\label{scattering}

In order to consider the scattering solutions for $x<x_0$, we solve the differential equation 

\begin{equation}
\label{x<x0}
\frac{d^2\phi(x)}{dx^2}+\left[\left(E+a \right)^2-m^2 \right]\phi(x)=0,
\end{equation}

Eq. (\ref{x<x0}) has the general solution 

\begin{equation}
\label{phi_I}
\phi_{\textnormal{I}}(x)=b_1\,e^{-irx}+c_1\,e^{irx},
\end{equation}
where $r=\sqrt{\left(E+a\right)^2-m^2}$.

\bigskip
The scattering solutions for $x_0<x<x_1$ are obtained by solving the differential equation

\begin{equation}
\label{x<x0andx1}
\frac{d^2\phi(x)}{dx^2}+\left[\left(E-a \right)^2-m^2 \right]\phi(x)=0,
\end{equation}

Eq. (\ref{x<x0andx1}) has the general solution 

\begin{equation}
\label{phi_II}
\phi_{\textnormal{II}}(x)=b_2\,e^{-iqx}+c_2\,e^{iqx},
\end{equation}
where $q=\sqrt{\left(E-a\right)^2-m^2}$.

\bigskip
Now we consider the solutions for $x>x_1$

\begin{equation}
\label{eq_x}
\frac{d^2\phi(x)}{dx^2}+\left\{\left[E-a\tanh(bx)\right]^2-m^2\right\}\phi(x)=0.
\end{equation}

\medskip
On making the substitution $y=-e^{2bx}$ , Eq. (\ref{eq_x}) becomes

\begin{equation}
\label{eq_y)}
4b^2 y \frac{d}{dy}\left[y\frac{d\phi(y)}{dy}\right]+\left[\left(E+a\frac{1+y}{1-y} \right)^2-m^2 \right]\phi(y)=0.
\end{equation}

\medskip
Putting $\phi_(y)=y^\alpha(1-y)^\beta f(y)$, Eq. (\ref{eq_y)}) reduces to the hypergeometric differential equation

\begin{equation}
\label{eq_hyper}
y(1-y)f''+[(1+2\alpha)-(2\alpha+2\beta+1)y]f'-(\alpha+\beta-\gamma)(\alpha+\beta+\gamma)f=0,
\end{equation}
where the primes denote derivates with respect to $y$ and the parameters $\alpha$, $\beta$, and $\gamma$ are

\begin{eqnarray}
\label{alpha}
\alpha&=&i\nu \,\,\, \textnormal{with} \,\,\, \nu=\frac{\sqrt{(E+a)^2-m^2}}{2b},\\
\label{beta}
\beta&=&\lambda \,\,\, \textnormal{with} \,\,\, \lambda=\frac{b+\sqrt{b^2-4a^2}}{2b},\\
\label{gamma}
\gamma&=&i\mu \,\,\, \textnormal{with} \,\,\, \mu=\frac{\sqrt{(E-a)^2-m^2}}{2b}.
\end{eqnarray}

\medskip
Eq. (\ref{eq_hyper}) has the general solution in terms of Gauss hypergeometric functions $_2F_1(\mu,\nu,\lambda;y)$ \cite{abramowitz:1965}

\begin{eqnarray}
\label{sol_y}
\nonumber
 \phi(y)&=&C_1 y^\alpha \left(1-y\right)^\beta\, _2F_1\left(\alpha+\beta-\gamma,\alpha+\beta+\gamma,1+2\alpha;y\right)\\
 &+&C_2 y^{-\alpha} \left(1-y\right)^\beta\, _2F_1\left(-\alpha+\beta-\gamma,-\alpha+\beta+\gamma,1-2\alpha;y\right).
\end{eqnarray}

\medskip
In terms of variable $x$ Eq. (\ref{sol_y}) becomes

\begin{eqnarray}
\label{sol_x}
\nonumber
\phi(x)&=&C_1 \left(-e^{2bx}\right)^{i\nu} \left(1+e^{2bx}\right)^\lambda\, _2F_1\left(i\nu+\lambda-i\mu,i\nu+\lambda+i\mu,1+2 i\nu;-e^{2bx}\right)\\
\nonumber
 &+&C_2 \left(-e^{2bx}\right)^{-i\nu} \left(1+e^{2bx}\right)^\lambda\, _2F_1\left(-i\nu+\lambda+i\mu,-i\nu+\lambda-i\mu,1-2 i\nu;-e^{2bx}\right).\\
\end{eqnarray}

\medskip
From Eq. (\ref{sol_x}) the reflected and incident waves are

\begin{equation}
\label{phi_ref}
\phi_{\textnormal{ref}}(x)=b_3\,\left(1+e^{2bx}\right)^\lambda e^{-2ib\nu x}\, _2F_1\left(-i\nu+\lambda+i\mu,-i\nu+\lambda-i\mu,1-2 i\nu;-e^{2bx}\right).
\end{equation}

\begin{equation}
\label{phi_inc}
\phi_{\textnormal{inc}}(x)=c_3\,\left(1+e^{2bx}\right)^\lambda e^{2ib\nu x}\, _2F_1\left(i\nu+\lambda-i\mu,i\nu+\lambda+i\mu,1+2 i\nu;-e^{2bx}\right).
\end{equation}

\medskip
We define

\begin{equation}
 \label{phi_III}
 \phi_{\textnormal{III}}(x)=b_3\,\phi_{\textnormal{ref}}(x)+c_3\,\phi_{\textnormal{inc}}(x) 
\end{equation}

Using the relation \cite{abramowitz:1965} 

\begin{eqnarray}
 \label{relation}
 \nonumber
 _2F_1(a,b,c;z)&=&\frac{\Gamma(c)\Gamma(b-a)}{\Gamma(b)\Gamma(c-a)}(-z)^{(-a)}\,_2F_1(a,1-c+a,1-b+a;z^{-1})\\
 &+&\frac{\Gamma(c)\Gamma(a-b)}{\Gamma(a)\Gamma(c-b)}(-z)^{(-b)}\,_2F_1(b,1-c+b,1-a+b;z^{-1}),
\end{eqnarray}
with Eqs. (\ref{phi_ref}), (\ref{phi_inc}) and keeping the solution that asymptotically corresponds to a transmitted particle moved from left to right, the transmited wave becomes

\begin{equation}
\label{phi_trans}
\phi_{\textnormal{trans}}(x)=c_4\,e^{-2b\lambda x} \left(1+e^{2bx}\right)^\lambda e^{2ib\mu x}\, _2F_1\left(i\nu+\lambda-i\mu,-i\nu+\lambda-i\mu,1-2 i\mu;-e^{-2bx}\right).
\end{equation}

We define

\begin{equation}
\label{phi_IV}
\phi_{\textnormal{IV}}(x)=\phi_{\textnormal{trans}}(x).
\end{equation}

\medskip
When $x\rightarrow \pm \infty$ the $V \rightarrow \pm a$ and the asymptotic behaviour of Eqs. (\ref{phi_I}) and,  (\ref{phi_IV}) are plane waves with the relations of dispersion $r$, $q$, $\nu$ and, $\mu$

\begin{eqnarray}
\label{phi_asym}
\phi_{\textnormal{I}}(x)&=& b_1\,e^{-irx}+c_1\,e^{irx},\\
\phi_{\textnormal{IV}}(x)&=& c_4\,e^{2ib\mu x},
\end{eqnarray}

\medskip
In order to find $R$ and $T$, we used the definition of the electrical current density for the one-dimensional Klein-Gordon equation (\ref{klein})

\begin{equation}
\label{current}
\vec{j}=\frac{i}{2}\left(\phi^*\vec{\nabla}\phi-\phi\vec{\nabla}\phi^*\right)
\end{equation}

\medskip
The current as $x \rightarrow -\infty$ can be descomposed as $j_\textnormal{L}=j_\textnormal{I}$, where $j_\textnormal{I}$ correspond to the incident current $j_\textnormal{inc}$ minus the reflected current $j_\textnormal{ref}$. Analogously we have that, on ther right side, as $x \rightarrow \infty$ the current is $j_\textnormal{R}=j_\textnormal{IV}$, where $j_\textnormal{IV}$ is the transmitted current $j_\textnormal{tras}$\cite{rojas:2005}.

\medskip
The reflection coefficient $R$, and the trasmission coefficient $T$, in terms of the reflected $j_\textnormal{inc}$, $j_\textnormal{ref}$, and $j_\textnormal{trans}$ currents are

\begin{equation}
\label{R}
R=\frac{j_\textnormal{ref}}{j_\textnormal{inc}}=\frac{|b_1|^2}{|c_1|^2}.
\end{equation}

\begin{equation}
\label{T}
T=\frac{j_\textnormal{trans}}{j_\textnormal{inc}}=\frac{2b\mu}{r}\frac{|c_4|^2}{|c_1|^2}.
\end{equation}

\medskip
The reflection coefficient $R$, and the transmission coefficient $T$ satisfy the unitary relation $T+R=1$ and are expresses in terms of the coefficients $b_1$, $c_1$ and, $c_4$. 
In order tor obtain $R$ and $T$ we proceed to equate at $x=x_1$, $x=x_0$ and, $x=0$ the wave functions and their first derivatives. From the matching condition we derive a system of equations governing the dependende of coefficients $b_1$, $b_2$, $c_1$, $c_2$, $b_3$ and,  $c_3$ on $c_4$ that can solve numerically with the Software Wolfram {\it Mathematica} 9. 

The dispersion relations $r$ and $\mu$ must be positive because it correspond to an incident particle moved from left to right and,  their sign depends on the group velocity, defined by \cite{calogeracos:1999}

\begin{equation}
 \label{group_r}
 \frac{dE}{dr'}=\frac{r'}{E+a}\geq 0.
\end{equation}

\begin{equation}
 \label{group_mu}
 \frac{dE}{d\mu'}=\frac{\mu'}{E-a}\geq 0.
\end{equation}

\medskip
For the step potential with hyperbolic tangent potential we have studied three different regions. In the region $a-m>E>m$, $\mu'<0$ and, $r'>0$ we have that $R>1$, so superradiance occurs. In the region $a+m>E>a-m$ the dispersion relations $\mu$ and $r$ are imaginary pure and the transmited wave is attenuated, so $R=1$. In the region $E>a+m$ we observed transmission resonances. 

\begin{figure}[!th]
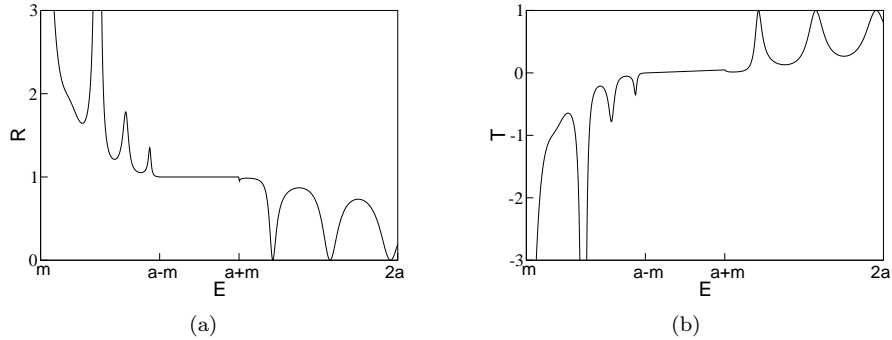

\bigskip
\begin{center}
\subfigure[]{\includegraphics[scale=0.22]{R_a=5,b=2_1.eps}}
\hspace{1cm}\subfigure[]{\includegraphics[scale=0.22]{T_a=5,b=2_1.eps}}
\end{center}
\caption{\label{TandR_b=2_1} The reflection $R$ and transmission $T$ coefficients varying energy $E$ for the step potential with hyperbolic tangent potential for $m=1$, $a=5$, $b=2$, $x_0=-4$ and, $x_1=-2$.}
\end{figure}

\begin{figure}[!th]
\begin{center}
\subfigure[]{\includegraphics[scale=0.22]{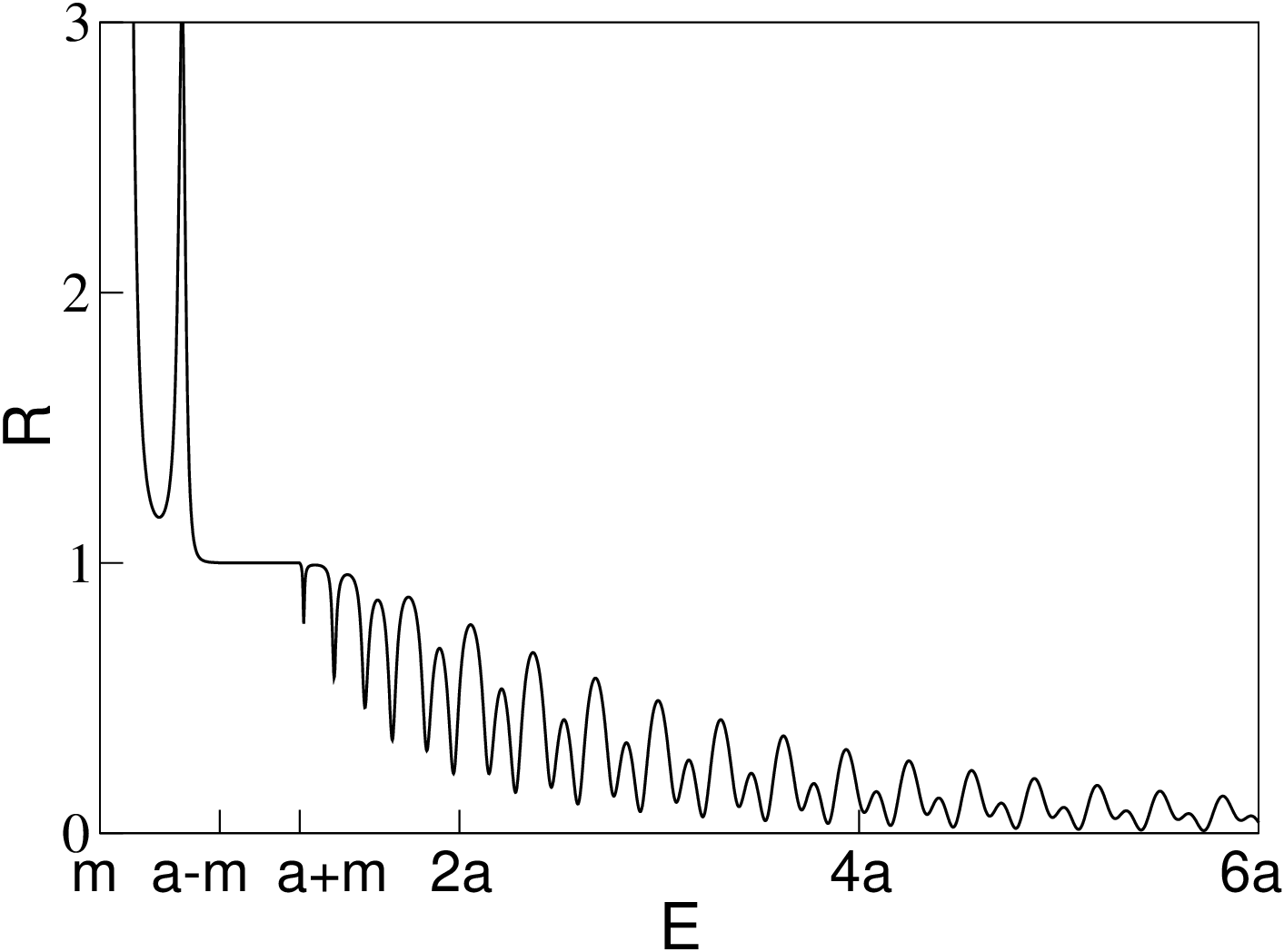}}
\hspace{1cm}\subfigure[]{\includegraphics[scale=0.22]{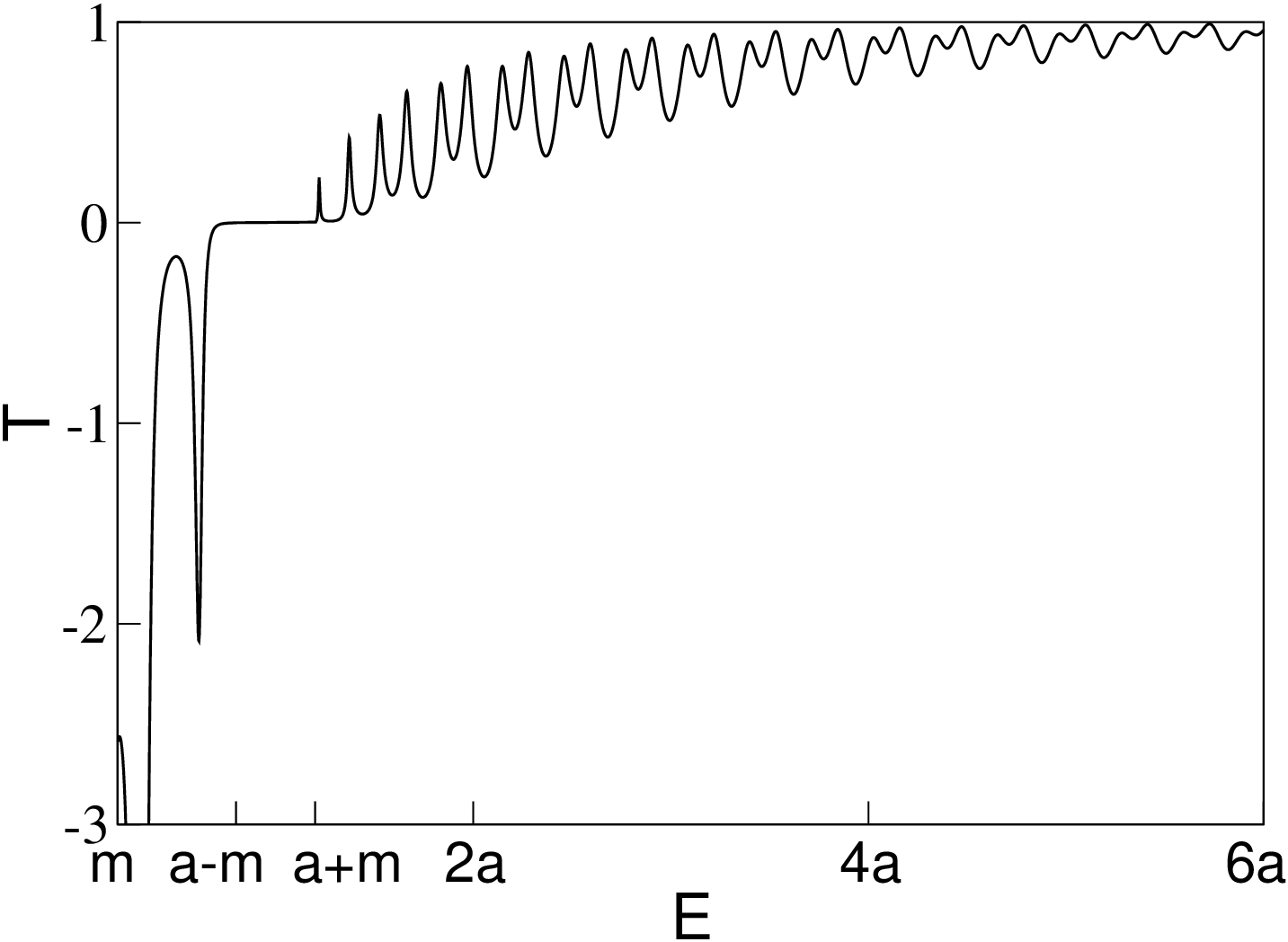}}
\end{center}
\caption{\label{TandR_b=50_1} The reflection $R$ and transmission $T$ coefficients varying energy $E$ for the step potential with hyperbolic tangent potential for $m=1$, $a=5$, $b=50$, $x_0=-4$ and, $x_1=-2$.}
\end{figure}

\begin{figure}[!th]
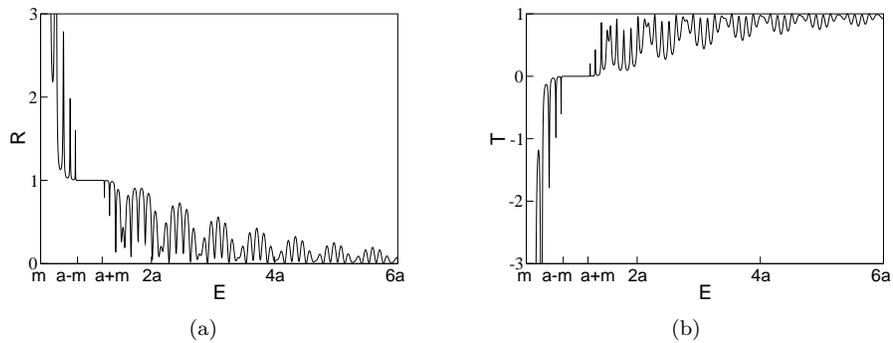

\begin{center}
\subfigure[]{\includegraphics[scale=0.22]{R_a=5,b=50_2.eps}}
\hspace{1cm}\subfigure[]{\includegraphics[scale=0.22]{T_a=5,b=50_2.eps}}
\end{center}
\caption{\label{TandR_b=50_2} The reflection $R$ and transmission $T$ coefficients varying energy $E$ for the step potential with hyperbolic tangent potential for $m=1$, $a=5$, $b=50$, $x_0=-6$ and, $x_1=-1$.}
\end{figure}

\bigskip
Figs. \ref{TandR_b=2_1}(a) and \ref{TandR_b=2_1}(b)  shows the  reflection $R$ and transmission $T$ coefficients for $E>m$, $a=5$, $b=2$, $x_0=-4$ and, $x_1=-2$. Fig. \ref{TandR_b=50_1}(a) and \ref{TandR_b=50_1}(b) shows the  reflection $R$ and transmission $T$ coefficients $R$ for $E> m$, $a=5$, $b=50$, $x_0=-4$ and, $x_1=-2$. Figs. \ref{TandR_b=2_2}(a) and \ref{TandR_b=2_2}(b)  shows the  reflection $R$ and transmission $T$ coefficients for $E>m$, $a=5$, $b=2$, $x_0=-6$ and, $x_1=-1$. Fig. \ref{TandR_b=50_2}(a) and \ref{TandR_b=50_2}(b) shows the  reflection $R$ and transmission $T$ coefficients $R$ for $E> m$, $a=5$, $b=50$, $x_0=-6$ and, $x_1=-1$. 
We observed in Figs. \ref{TandR_b=2_1}(a), \ref{TandR_b=50_1}(a), \ref{TandR_b=2_2}(a) and, \ref{TandR_b=50_2}(a) that in the region $a-m>E>m$ the reflection coefficient $R$ is bigger than one whereas the coefficient of transmission $T$ becomes negative, so we observed superradiance \cite{cheng:2009, wagner:2010}. In Figs. \ref{TandR_b=2_1}(b), \ref{TandR_b=50_1}(b), \ref{TandR_b=2_2}(b) and, \ref{TandR_b=50_2}(b)  we find transmission resonances in the region $E>a+m$. On the other hand, the change in the values ​of ​$x_0$ and $x_1$ affects both the phenomenon of superradiance and transmission resonances: the position,  height and numbers of the peaks depends on $x_0$ and $x_1$.

\begin{figure}[!th]
\begin{center}
\subfigure[]{\includegraphics[scale=0.22]{R_a=5,b=2_2.eps}}
\hspace{1cm}\subfigure[]{\includegraphics[scale=0.22]{T_a=5,b=2_2.eps}}
\end{center}
\caption{\label{TandR_b=2_2} The reflection $R$ and transmission $T$ coefficients varying energy $E$ for the step potential with hyperbolic tangent potential for $m=1$, $a=5$, $b=2$, $x_0=-6$ and, $x_1=-1$.}
\end{figure}

%Conclusion
\section{Conclusion}
\label{conclusion}
In this paper we have discussed the scattering solutions of the Klein-Gordon equation for a step potential with hyperbolic tangent potential.
The solutions are determined in terms of hypergeometric functions. The calculation of the reflection coefficient $R$ and transmission coefficient $T$ is shown. For the region where $a-m>E>m$, the phenomenon of superradiance occurs. In the case $b=2$  transmission resonances are observed for $a+m<E<2a$ in the two cases considered, while in the case where $b=50$ transmission resonances appears for $E>4a$ for $x_0=-4$, $x_1=-2$ and $E>2a$ for $x_0=-6$, $x_1=-1$.  

%Bibliography
%\bibliography{btanh}

\end{document}